\documentclass[twocolumn,prp,aps]{revtex4}
\usepackage{amsmath}
\usepackage{amssymb}
\usepackage{graphicx}
\usepackage{dcolumn}
\usepackage{graphicx}
\usepackage{dcolumn}
\usepackage{bm}

\begin{document}

%\preprint{Phys.Rev.Lett.}

\title{Persistence of two-dimensional topological insulator state in wide HgTe quantum well}
\author{E. B. Olshanetsky,$^1$ Z. D. Kvon,$^{1,2}$ G. M. Gusev,$^3$  A. D. Levin,$^3$ O. E. Raichev,$^4$ N. N. Mikhailov,$^1$ and S. A. Dvoretsky,$^{1}$}

\affiliation{$^1$Institute of Semiconductor Physics, Novosibirsk
630090, Russia}

\affiliation{$^2$Novosibirsk State University, Novosibirsk 630090,
Russia}

\affiliation{$^3$Instituto de F\'{\i}sica da Universidade de S\~ao
Paulo, 135960-170, S\~ao Paulo, SP, Brazil}

\affiliation{$^4$Institute of Semiconductor Physics, NAS of
Ukraine, Prospekt Nauki 41, 03028 Kyiv, Ukraine}

\date{\today}
\begin{abstract}
Our experimental studies of electron transport in wide (14 nm) HgTe
quantum wells confirm persistence of a two-dimensional topological
insulator state reported previously for narrower wells, where it was
justified theoretically. Comparison of local and nonlocal resistance
measurements indicate edge state transport in the samples of about 1
mm size at temperatures below 1 K. Temperature dependence of the
resistances suggests an insulating gap of the order of a few meV. In
samples with sizes smaller than 10 $\mu$m a quasiballistic transport
via the edge states is observed.


\end{abstract}

\maketitle

The topological insulators (TI) represent a quantum state of condensed
matter with insulating bulk and conducting gapless states at the surface
or edge \cite{hasan, qi, moore, moore2}. The existence of such materials
is justified within a concept of topological ordering introducing order
parameters which are often expressed as invariant integrals over the momentum
space. In the presence of time reversal symmetry, the materials with energy
band gaps (band insulators) are classified by $Z_2$ topological invariants
\cite{kanemele} which take two values, 1 or 0, thereby providing a
distinction between topological and normal insulators. Mathematically,
one can construct $Z_2$ invariants in different ways, but their physical
meaning always relies on the symmetry of electron wave function which is
changed as a result of energy band inversion. Such an inversion occurs due
to spin-orbit coupling and Darwin term contributions in the Hamiltonians
of the crystals formed from heavy atoms. There are three types of band
inversions ($s-p$, $p-p$, and $d-f$) in the three-dimensional (3D) TI
discovered so far \cite{zhang}.

The most extensively studied TI materials, bismuth chalcogenides
and related alloys, belong to $p-p$ inversion type. For thin
layers of these materials, one expects a dimensionality crossover:
when layer thickness $d$ decreases, the material transforms from
3D TI into two-dimensional (2D) TI \cite{liu}. This occurs when
the wave function decay length of the surface states becomes
comparable to $d$. As a result, the 2D states from opposite
surfaces hybridize and their spectrum is no longer gapless
\cite{liu,lu}. On the other hand, since the surface states in TI
cover the whole surface of the layer, including side regions, they
are transformed into 1D conducting edge channels in these regions,
Fig. 1 (a), \cite{wu}.

\begin{figure}[ht!]
\includegraphics[width=0.9\linewidth]{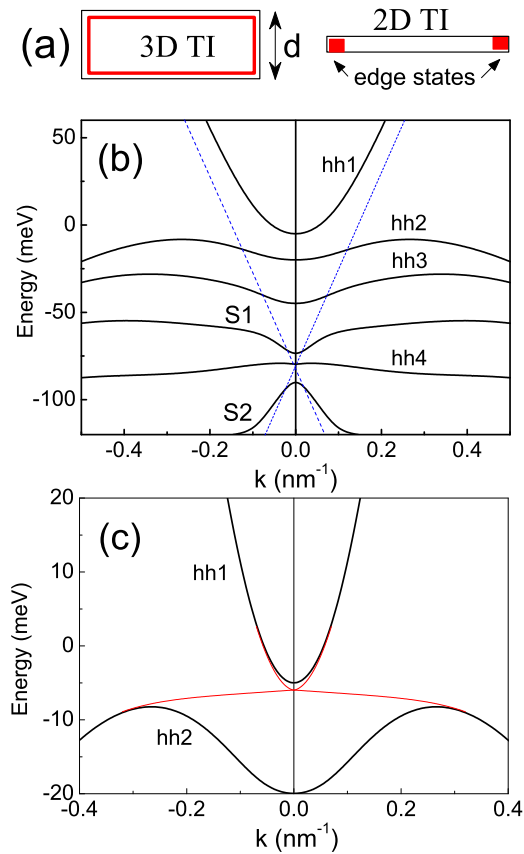}
\caption{\label{fig.1}(Color online) (a) Cross-section of a 3D TI
sample (schematic) with surface state channels shown in red. As the
thickness of the sample decreases, a transition to 2D TI takes
place. (b) Energy spectrum of size-quantized subbands in a symmetric
14 nm wide Cd$_{0.65}$Hg$_{0.35}$Te/HgTe/Cd$_{0.65}$Hg$_{0.35}$Te
QW, calculated numerically by using Kane Hamiltonian. Dashed (blue)
lines show the spectrum of interface states at a single
HgTe/Cd$_{0.65}$Hg$_{0.35}$Te boundary under approximation that
mixing of these states with $hh$ states is neglected. (c) A
magnified picture of two upper $hh$ subbands and expected edge state
spectrum, thin (red) lines, in the gap between them.}
\end{figure}

A special place in this connection belongs to HgTe, a
zinc-blend-structure crystal with $s-p$ band inversion, where the
energy of $p$-type $\Gamma_8$ band in the $\Gamma$ point of
Brillouin zone is higher than the energy of $s$-type $\Gamma_6$
band. In spite of the inversion, bulk HgTe is not a 3D TI because
it is a symmetry-protected gapless material. A gap can be opened
in a thin HgTe layer sandwiched between Cd$_{1-x}$Hg$_x$Te layers
(normal insulators), as realized in epitaxially grown quantum
wells (QWs). Due to size quantization, the heavy hole ($hh$)
continuum that forms the valence band in HgTe splits into a set of
2D states hybridized with two interface-like states $S1$ and $S2$,
Fig. 1 (b). The gap between the ground-state $hh$ subband ($hh1$)
and the next adjacent subband exists for QW narrower than $18$ nm
(for wider wells the QW is in a semimetallic state). Since the gap
opening is accompanied with the dimensionality crossover, a
HgTe-based QW should be a 2D TI having edge states in the gap
between subbands. However, a direct theoretical proof for this
statement, based on a two-subband effective 2D Hamiltonian
\cite{bernevig}, has been done only for a special situation when
$S1$ subband is just slightly below $hh1$ subband, which is
applicable to narrow wells in the width range of approximately
$6.3-8.3$ nm. The edge state transport in such QWs was also
confirmed experimentally \cite{konig}. In wider HgTe QWs, any
effective Hamiltonian methods are not generally feasible because
of a complicated subband structure, though a usage of
three-subband ($S1$, $hh1$, and $hh2$) basis \cite{raichev}
extends the range of applicability of such methods. In particular,
it was found that when $S1$ subband falls below $hh2$ one (so the
principal gap is formed between $hh1$ and $hh2$ subbands) the edge
states exist both in this gap and in the next gap between $hh2$
and $S1$ subbands.

A question that naturally arises concerns persistence of 2D TI state
in wide QWs where the $S1$ subband lies below several $hh$ subbands,
so the situation is far different from that described theoretically
in Refs. \cite{bernevig} and \cite{raichev}. From the point of view of
dimensionality crossover, there are no reasons to
deny the 2D TI nature of these systems, since widening of the QW
(actually, approaching of HgTe layer to the bulk state) does not
cancel the fact of $s-p$ inversion in this layer and, accordingly,
cannot destroy the edge states. In this Letter we report
experimental investigation of 14 nm wide HgTe QWs which are wider
than those studied previously \cite{konig, roth, gusev} but still
have a sizeable gap of a few meV between $hh1$ and $hh2$ subbands.
We plot the expected edge states in this gap schematically in Fig. 1
(c) as two (one for each spin number) gapless branches merging with
2D subbands on a tangent. By transport measurements, we indeed
obtain numerous proofs for the edge state transport in these QWs.

\begin{figure}[ht!]
\includegraphics[width=0.9\linewidth]{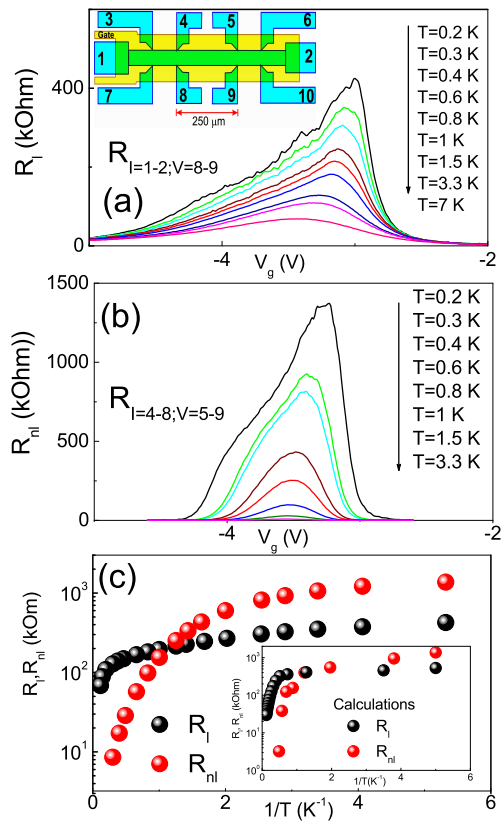}
\caption{\label{fig.2}(Color online) Local (a) and nonlocal (b)
resistance at different temperatures, sample I, $B=0$
(insert - schematic view of the sample). (c) Temperature dependence of
local and nonlocal resistance when the Fermi energy is situated in
the center of the insulating bulk gap (insert - calculation).}
\end{figure}

The experimental samples are Hall bridges fabricated on top of the
14 nm wide HgTe QW with the surface orientation (112) and provided
with electrostatic gate. Their fabrication technology is described
in detail in \cite{kvon2}. Three different types of experimental
samples were used: macroscopic Hall bridges (see Fig. 2) with the
width 50 $\mu$m and the distance between the voltage probes 100
$\mu$m and 250 $\mu$m, and two  types of microscopic samples,
whose layout together with the scale is shown in Fig. 3
\cite{supplemental}. The transport measurements were conducted in
the temperature range $0.2-10$ K and in magnetic fields up to 10 T
using the standard phase detecting scheme on frequencies $3 - 12$
Hz and the driving current $0.1 - 1$ nA to avoid heating effects.
The electron mobility $\mu$ in all samples studied was above
$10^5$ cm$^2$/V s for the carrier density $3\times10^{11}$
cm$^{-2}$.

We first consider the properties of our macroscopic samples.
Observation of both local ($R_{l}$) and nonlocal ($R_{nl}$)
resistance in the absence of magnetic field is generally
considered as a direct proof of 2D TI state. In that case the
current flows through a sample along its borders and resistance
exists regardless to positions of voltage probes with respect to
current contacts, either in ballistic \cite{roth} or diffusive
\cite{gusev} regime. Figures 2 (a) and 2 (b) show $R_{l}$ and
$R_{nl}$ as functions of the gate voltage for various temperatures
in the range $0.2 - 3.3$ K and Fig. 2 (c) shows these resistances
as functions of temperature for the gate voltages corresponding to
the maxima of local and nonlocal resistance, i.e. when the Fermi
level is in the middle of the insulating gap. Qualitatively, the
behavior of local and nonlocal resistances is similar. When the
temperature is above 1 K, both resistances grow exponentially with
decreasing temperature. For lower temperatures the resistance
growth slows down and for $T < 0.5$ K it obeys a power law $R
\propto T^{-\alpha}$ $(\alpha \approx 0.5)$. As this takes place,
even though at higher temperatures the local resistance is several
orders of magnitude greater than the nonlocal one, at $T < 1$ K
the nonlocal resistance becomes larger. The described behavior of
$R_{l}$ and $R_{nl}$ is typical for the 2D TI, as it follows from
the fundamental difference in the relative contributions of the
bulk and edge transport when measuring in local and nonlocal
configurations \cite{gusev}. At higher temperatures, when the bulk
contribution is still sufficiently large, $R_{nl}$ is
exponentially small compared to the sheet resistivity. With
decreasing $T$ the bulk contribution to transport also decreases,
and at some $T$, depending on the insulating bulk gap, becomes
negligibly small. Under these conditions the difference in the
resistance values measured in local and nonlocal configurations is
determined only by the distribution of currents flowing along the
sample perimeter and by the position of the voltage probes. At $T
< 0.5$ K for all investigated configurations the resistance at its
maximum is more than an order of magnitude greater than $h/2e^2$
which means that the transport via the edge states is diffusive.

Let us discuss in more detail the temperature dependence of
$R_{nl}$ and $R_{l}$. First, the maximum of the curves
$R_{nl}(V_g)$ and $R_{l}(V_g)$ shifts to the right with the
temperature decreasing. Such behavior was not observed in 2D TI
previously and is probably related to the complicated energy
spectrum of the system investigated, or, more specifically to its
much smaller gap which is further diminished by the bulk bands
density of states tails. Second, as has already been mentioned,
with lowering temperature the resistance, after the exponential
growth, continues to increase, but at a much lower rate, following
a power law $R \propto T^{0.5}$, typical for quasi 1D wires in a
weak localization regime. We have attempted a quantitative
description of the temperature dependence of the local and
nonlocal resistance peak values by using the model proposed in
\cite{gusev3, supplemental}. The activation energy for the bulk
transport has been chosen as a fitting parameter. The results of
the calculation presented in Fig. 2 (c) show a reasonable
agreement between the calculated and measured dependences. The
value of the activation energy found from fitting the calculation
to experiment is approximately 1.2 meV for local and nonlocal
configurations alike, as was expected. This value is less than
$\Delta$= 3.3 meV, the indirect insulating gap value obtained from
the energy spectrum calculation. This discrepancy is not
surprising if one considers the disorder due to impurities and QW
thickness fluctuations that is always present in a real HgTe
sample. In such case one may expect that the activation energy
would correspond to the mobility gap rather than to a much larger
calculated insulating bulk gap \cite{Prange}.

 The mean free path for the edge states transport determined experimentally for the diffusive transport in the macroscopic samples is $2-5$ $\mu$m for
sample I and $12-14$ $\mu$m for sample II (See supplementary
material). On this account our QWs look promising for observation
of ballistic transport via edge states, considering that in the
majority of the previously studied 2D TI this value was close to 1
$\mu$m. For this purpose, two types of microscopic samples were
fabricated, one with the dimensions $W=1.7$ $\mu$m and $L=1.8$
$\mu$m (see Insert to Fig.3a) and the other an H-shaped bridge
with the width $W=3.2$ $\mu$m and the length $L=2.8$ $\mu$m (see
Insert to Fig.3b).

\begin{figure}[ht!]
\includegraphics[width=0.9\linewidth]{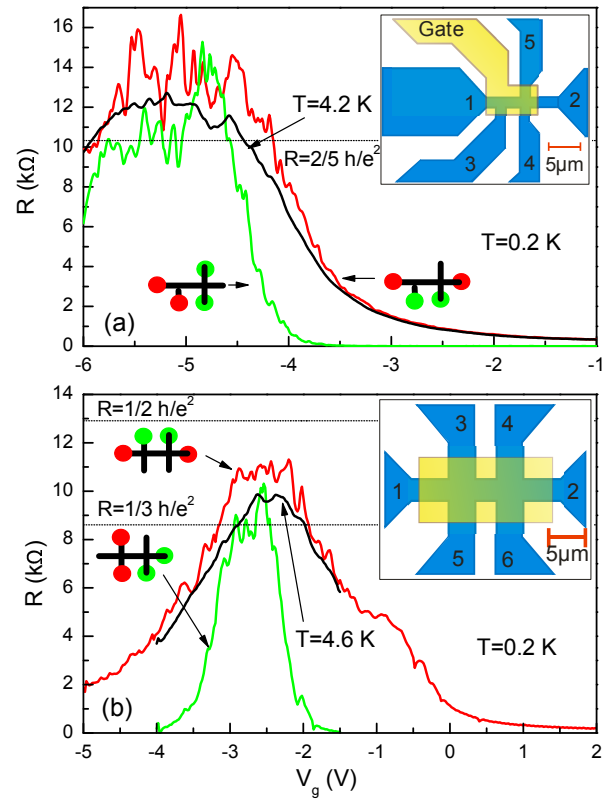}
\caption{\label{fig.3}(Color online) $R(V_g)$ dependences for
local (red curves) and nonlocal (green curves) resistance
measurement configurations obtained in two different types of
microscopic samples. The samples layouts are shown schematically
in the Inserts. All curves were obtained at 0.2 K except the black
curves measured at temperatures 4.2 and 4.6 K. Opposite each curve
the corresponding measurement configuration is shown
schematically. The dashed horizontal lines mark the resistance
values expected for these configurations in the case of a purely
ballistic transport via edge current states.}
\end{figure}

Figure 3 shows the dependence of the local (red color) and
nonlocal (green color) resistance versus gate voltage measured in
the two microscopic samples. When the bias applied to the gate
decreases below $-1$ V (Fig.3a) ($2$ V in Fig.3b), the local
resistance starts to grow gradually as the Fermi level first
descends to the bottom of the conduction band and then enters the
insulating bulk gap at $V_g \approx -4.5$ V (Fig.3a) ($V_g \approx
-2$ V in Fig.3b). The nonlocal response is close to zero when the
Fermi level remains in the conduction band. The abrupt increase in
the nonlocal resistance as the Fermi level enters the insulating
gap signals, as well as in the macroscopic case, the presence of
the edge current states in the gap. The calculation of the local
and nonlocal resistance values expected in our microstructures in
the case of ballistic transport via edge states is quite simple
and is indicated by the dashed lines for each configuration shown
in Fig.3.  As may be seen, the average resistance values measured
at $T=0.2$ K in the gate voltage range corresponding to the Fermi
level staying in the insulating gap are quite close to the levels
expected for purely ballistic transport. With the negative bias
increasing the Fermi level enters the valence band and the sample
resistance decreases. Relying on this data, it is possible to
conclude that our microstructures demonstrate a quasiballistic
edge transport in a 2D TI. This fact is quite important
considering that up till now the observations of ballistic
transport in HgTe-based 2D TI reported in \cite{konig} has
remained unique. We also observe other similarities with the
results reported in \cite{konig}. In particular, when the Fermi
level lies in the insulating gap there are random fluctuations
both in the local and nonlocal resistance. The amplitude of these
fluctuations sharply decreases as the temperature increases
indicating their mesoscopic origin.

Figure 3 also shows the variation of the local resistance with
temperature in the range from 0.2 K to 4.2K in our
microstructures. One can see that this variation is noticeably
weaker than in our macroscopic samples (compare with Fig. 2 (c)).
This fact has a simple explanation. In microstructures the sheet
conductance due to the bulk transport is the same as in
macroscopic samples while the resistance to transport via edge
states decreases by approximately one order of magnitude. This
observation also indicates that the mean free path for the edge
states transport in our QWs must be comparable to or higher than
$\approx 10$ $\mu$m. We also observe a suppression of both local
and nonlocal quasiballistic conductance by weak magnetic field, a
behavior typical for 2D TI state \cite{konig, gusev2}, which will
be reported elsewhere.

We believe that our observation of relatively high mean free path
for the edge states is not accidental but rather related to the
advantages associated with the use of a wider quantum well.
Indeed, the width of any QW is not uniform but fluctuates from
point to point around its average value $d$ with the amplitude
$\delta$. That $\delta$ is determined by the growth technology
employed and is practically independent of the QW width. The
fluctuation of the QW width results in a random potential in the
bulk of the QW. However, the amplitude of  that random potential
would be much smaller in a wider QW as it is proportional to
$1/(d^3)$. From that point of view it is clear that a wider QW
well is more advantageous for the observation of ballistic
transport in 2D TI.

The results of the present study confirm that the 2D TI state in
HgTe QWs is quite robust and exists in a sizeable range of well
widths despite of the fact that the energy spectrum in such QWs is
complicated and strongly dependent on the well width.

The work was supported by the RFBI grants, by the RAS grant 24.11
and by FAPESP, CNPq (Brazilian agencies).

\end{document}

\preprint{Phys.Rev.Lett.}

\title{Online supplemental material to : Persistence of two-dimensional topological insulator state in wide HgTe quantum well}
\author{E. B. Olshanetsky,$^1$ Z. D. Kvon,$^{1,2}$ G. M. Gusev,$^3$  A. D. Levin,$^3$ O. E. Raichev,$^4$ N. N. Mikhailov,$^1$ and S. A. Dvoretsky,$^{1}$}

\affiliation{$^1$Institute of Semiconductor Physics, Novosibirsk
630090, Russia}

\affiliation{$^2$Novosibirsk State University, Novosibirsk 630090,
Russia}

\affiliation{$^3$Instituto de F\'{\i}sica da Universidade de S\~ao
Paulo, 135960-170, S\~ao Paulo, SP, Brazil}

\affiliation{$^4$Institute of Semiconductor Physics, NAS of Ukraine,
Prospekt Nauki 41, 03028 Kyiv, Ukraine}

\date{\today}

\begin{abstract}
In this supplementary, we provide details on the sample
description, the mean free path calculation and the model taking
into account the edge and bulk contribution to the total current.


\end{abstract}

\maketitle

\section{Sample description}
In the main text we describe the experimental results in
macroscopic and mesoscopic samples in zero magnetic field. The
Cd$_{0.65}$Hg$_{0.35}$Te/HgTe/Cd$_{0.65}$Hg$_{0.35}$Te quantum
wells with surface orientation (112) studied here were grown by
molecular beam epitaxy (MBE). The schematic view of the samples
layer structure is shown in Fig. 1a. As shown in previous
publications \cite{kvon}, the use of substrates inclined to the
singular orientations may lead to the growth of more perfect
films. Therefore, the efficient growth of alloys is done
predominantly on substrates with surface orientations of (112),
which deviate from the singular planes by approximately
$19^{\circ}$. We have investigated samples of both macroscopic and
microscopic dimensions. The microscopic devices were of two types.
The first with the dimensions $W=1.73$ $\mu$m and $L=1.8$ $\mu$m
and the other an H-type bridge with the width $W=3.2$ $\mu$m and
the length $L=2.8$ $\mu$m, Fig. 1 (b). The macroscopic sample
consists of 3 narrow (50 $\mu$m wide) consecutive segments of
different length (100, 250, 100 $\mu$m), and 8 voltage probes,
Fig.1(c). The ohmic contacts to the two-dimensional gas were
formed by the in-burning of indium. To prepare the gate, a
dielectric layer containing 100 nm SiO$_{2}$ and 100 nm
Si$_{3}$Ni$_{4}$ was first grown on the structure using the
plasmochemical method. Then, the TiAu gate was deposited.

\begin{figure}[ht!]
\includegraphics[width=9cm,clip=]{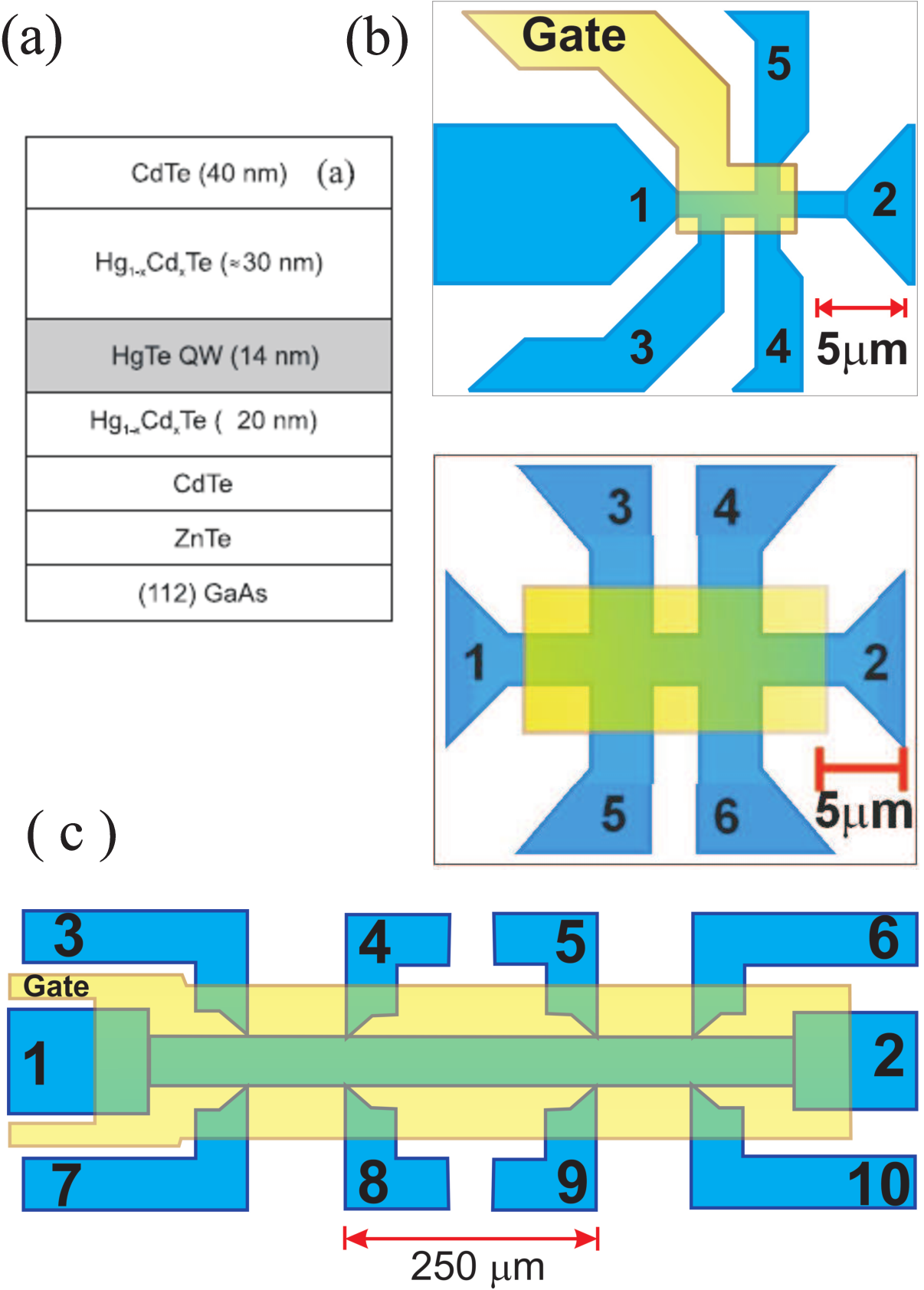}
\caption{\label{fig.1}(Color online) (a) Schematic layer structure
of a wafer containing 14 nm HgTe QW.  (b) mesoscopic samples
topology, yellow rectangular area depicts the gate; (c)
macroscopic samples layout.}
\end{figure}

Low-temperature magnetotransport measurements were carried out to
identify the charge neutrality point (CNP) and to obtain the
relation between the induced carrier density and the gate voltage
$V_{g}$. Sweeping the gate voltage, we have found a continuous
crossover from electron to hole type of conductivity, when the
Fermi level is shifted from the conduction to the valence band.
The possibility to tune the carrier density electrostatically
enables the observation of electron and hole transport in a single
device, making it possible to control the filling of Landau levels
in the presence of the perpendicular magnetic field. Figure 2
shows longitudinal $\rho_{xx}$ and Hall $\rho_{xy}$ resistivities
as functions of the gate voltage at a fixed magnetic filed $B=2$
T. Pronounced plateaux with values $\rho_{xy}=h/\nu e^{2}$ are
clearly seen for electrons at $\nu$ from $-1$ to $-5$, accompanied
by deep minima in $R_{xx}$. As $V_{g}$ is swept through the CNP,
when the Fermi level is in the middle of the gap, the longitudinal
resistivity shows a maximum, whereas $\rho_{xy}$ goes gradually
through zero from negative on the electron side to positive on the
hole side. It is a clear indication of CNP position and
electron-hole crossover as a function of the gate voltage.
Observation of the edge state transport near CNP in zero magnetic
field is described in the main text. From the resistance minima we
obtain the density variation with gate voltage $8.4 \times
10^{14}$ m$^{-2}$V$^{-1}$, which coincides with the density
extracted from the capacitance.

\begin{figure}[ht!]
\includegraphics[width=9cm,clip=]{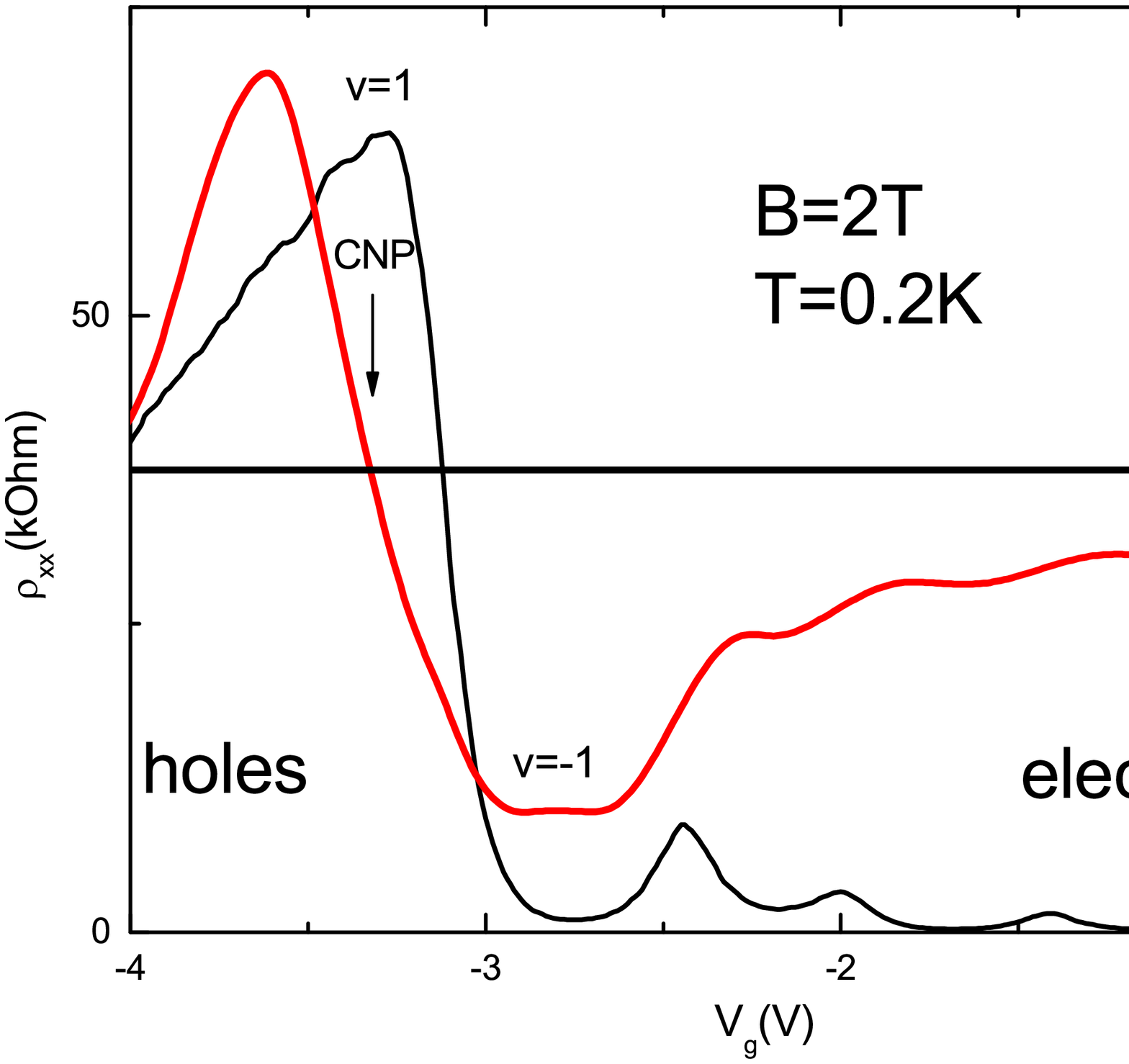}
\caption{\label{fig.2}(Color online) (a) Longitudinal $\rho_{xx}$
(black line) and Hall $\rho_{xy}$ (red line) resistivities as functions
of the gate voltage at a fixed magnetic field, $B=2.0$ T. Vertical arrow
indicates CNP.}
\end{figure}

\section{Mean free path calculation for the edge states transport}

At $T < 0.5$ K for all investigated measurement configurations in
the macroscopic sample the resistance at its maximum is more than
an order of magnitude greater than $h/2e^2$ which means that the
transport via the edge states is diffusive. In this case there is
a simple expression that allows one to calculate the resistance
value for any measurement configuration assuming that there is
only edge state transport in the sample:
$R_{n,m}^{i,j}=\frac{L_{n,m}L_{i,j}}{lL} (h/e^2)$, where
$R_{n,m}^{i,j}$ is the resistance measured between contacts $i$
and $j$ while the current is maintained between contacts $n$ and
$m$, $L_{i,j}$ ($L_{n,m}$) are the distances between $i$ and $j$
($n$ and $m$) along the sample edge that does not include $n$ and
$m$ ($i$ and $j$), $L$ is the total perimeter of the sample, and
$l$ is the mean free path. Thus, it is possible to determine the
mean free path $l$ for the transport via the edge states. Table 1
presents the results obtained in this way for samples I and II.
The values of $l$ derived from different measurement
configurations in the same sample are quite close, but may differ
substantially in different samples. This data supports the
statement advanced in \cite{gusev1} that in 2D TI fabricated on
the basis of HgTe QWs with inverted energy spectrum the edge
states have a macroscopic length (about a millimeter). On the
other hand, it is in agreement with a conclusion made earlier
\cite{konig,gusev1} concerning the absence of topological
protection against backscattering on the scale of few microns and
longer.

\begin{table}
\begin{tabular}{|c|c|c|}    \hline
Sample & Config. &  $l$ ($\mu$m) \\    \hline

I & $R_{1-2}^{8-9}$ & 4.6 \\    \hline

I & $R_{4-8}^{5-9}$ & 3.6 \\    \hline

I & $R_{1-2}^{7-8}$ & 2.6 \\    \hline

II & $R_{1-2}^{3-4}$ & 12 \\    \hline

II & $R_{1-2}^{7-8}$ & 12 \\    \hline

II & $R_{1-2}^{8-9}$ & 14 \\    \hline

II & $R_{5-9}^{4-8}$ & 13 \\    \hline

II & $R_{3-7}^{4-8}$ & 12.5 \\    \hline

II & $R_{3-7}^{5-9}$ & 11.6 \\    \hline

\end{tabular}
\caption{\label{Table 1.} The first column: sample number. The
second column: various local and nonlocal measurement
configurations. The third column: mean free path for the edge
states transport obtained from the comparison of the peak
resistance value with theory.}
\end{table}

\section{Edge+bulk model}

In the main text it has been demonstrated that the resistance of HgTe quantum wells
reveals a broad peak when the gate voltage induces an additional charge density
altering the quantum wells from n-type conductor to p-type one around
the charge neutrality point (CNP). The peak amplitude is approximately equal to
$h/2e^{2}$ in mesoscopic samples, due to nearly ballistic edge-state transport,
and $R \gg h/2e^{2}$ in macroscopic samples. The lack of the robustness of resistance
quantization against intrinsic and introduced disorder in macroscopic samples may be
explained in various ways \cite{maciejko,strom,vayrynen}. Independently of the
particular microscopic mechanism responsible for this, the resistance can be
described by the combination of the edge-state and bulk transport contributions, with
taking into account both the backscattering within one edge and bulk-edge coupling.
The local and nonlocal transport coefficients arise from the edge state contribution
and short-circuiting of the edge transport by the bulk contribution, the latter is
more important away from the CNP. Below we reproduce the basic features of this
model \cite{abanin, gusev}. The transport properties in the bulk can be described
by the current-potential relation

\begin{eqnarray}
%1
\textbf{j}_{i}({\bf r}) =-\hat{\sigma}_{i}\nabla\psi_{i}({\bf r}),
\end{eqnarray}
\[
\hat{\sigma}_{i} = \left( {\begin{array}{cc}
 \sigma^{(i)}_{xx} & \sigma^{(i)}_{xy} \\
 \sigma^{(i)}_{yx} & \sigma^{(i)}_{xx} \\
 \end{array} } \right),
\]
where $i=1,2$ numbers the states with different projections of the spin,
$\psi_i$ are the electrochemical potential for electrons, and ${\bf r}=(x,y)$
is the 2D coordinate. As we consider the isotropic conduction, the non-diagonal
part of the conductivity tensor appears only at non-zero magnetic field.
Assuming the components of the conductivity tensor as
coordinate-independent parameters, we can solve the problem by solving the
Laplace equation for the potentials, $\nabla^{2}\psi_{i}({\bf r})=0$, because
the charge conservation continuity conditions requires $\nabla\textbf{j}_{i}({\bf r})=0$.
A solution to Laplace equation is uniquely determined by specifying boundary
conditions, which in our case are modified by the bulk-edge current leakage. In
order to describe the transport in the presence of the edge states, we introduce
two phenomenological constants $\gamma$ and $g$, which represent edge to edge and
bulk to edge inverse scattering length, respectively. Then, the boundary conditions
expressing zero current normal to the boundary in the presence of the bulk-edge
coupling are given by
\begin{eqnarray}
%2
\textbf{n}\textbf{j}_{i} =g(\psi_{i}-\varphi_{i}),
\end{eqnarray}
where $\varphi_{i}$ are the local chemical potentials of the edge states, $\psi_{i}$ and
$\textbf{j}_{i}$ are the potentials and currents at the boundary, and $\textbf{n}$ is
a unit vector normal to the boundary.

The edge state transport can be described by continuity equations
\cite{abanin, dolgopolov} taking into account the scattering between
edge and bulk:
\begin{eqnarray}
%1
\partial_{x}\varphi_{1}=\gamma(\varphi_{2}-\varphi_{1})+g(\psi_{1}-\varphi_{1}),\\
-\partial_{x}\varphi_{2}=\gamma(\varphi_{1}-\varphi_{2})+g(\psi_{2}-\varphi_{2}).
\end{eqnarray}
The general solution of this problem, therefore, includes solution
of a 2D Laplace equation for the bulk electrochemical potentials $\psi_{1,2}(x,y)$
together with Eqs. (2),(3),(4) describing the scattering between edge states and
between edge and bulk states. The current can be calculated from this solution
as a sum of the contributions from bulk and edge states.

In nonlocal configurations the edge $+$ bulk model can be solved only
numerically. We have performed self-consistent calculations to
find $\psi_{1,2}$ solution of Laplace equation in two space dimensions
and $\varphi_{1,2}$ solutions of equations 3 and 4 on the edge using the
Hall bar geometry schematically shown in Fig. 1 (c) [or Fig. 2 of the main
text]. The contacts are assumed to be thermal reservoirs, where full mixing
of electron spin states and bulk states occurs \cite{dolgopolov}. Note that,
in contrast to the standard QHE, when mixing of the edge states occurs within
metallic Ohmic contacts, in our samples potential mixing is provided by 2D
electron gas in the region outside of the metallic gate.

The equations for $\psi_{1,2}$ are discretized by the finite element method.
The generalized Neumann boundary conditions, Eq. (2), are set in the regions
outside the metal contacts. To solve the boundary value problem for a system
of ordinary differential equations (3) and (4) we use a finite difference code
that implements the 3-stage Lobatto IIIa formula. The boundary conditions
inside the metal contacts are set to $\varphi_{1,2}=\psi_{1,2}$.

For both local and non-local configurations, the resistance is calculated as
\begin{eqnarray}
%5
R_{xx} =V\;I_{tot}^{-1},\;I_{tot}=I_{edge}+I_{bulk}, \nonumber \\[0.6\baselineskip]
V =\frac{1}{2}\left( \varphi _{11}-\varphi _{11^{\prime }}+\varphi
_{21}-\varphi _{21^{\prime }}\right) ,
\end{eqnarray}
where $V$ is the potential difference at the voltage probes, $I_{tot}$ is the total
current flowing between current contacts, $\varphi _{i1}$ and $\varphi _{i1^{\prime} }$
are the potentials at the voltage probe locations. The edge and bulk currents for the
local case (at arbitrary point $x$ along the sample) are given by
\begin{eqnarray}
%6
I_{edge}=\frac{e^2}{h} \left(\varphi_{1}-\varphi_{2}+\varphi_{2^{\prime }}-\varphi
_{1^{\prime }}\right), \nonumber \\
I_{bulk}=\sum_{i=1,2} \left[ \sigma_{xy}^{\left( i\right) }\left( \psi_{i}-\psi_{i^{\prime}} \right)
-\sigma_{xx}^{\left( i\right) } \int dy \frac{\partial \psi_{i}}{\partial x} \right],
\end{eqnarray}
where $\varphi_{i}$, $\psi_{i}$ and $\varphi_{i^{\prime }}$, $\psi_{i^{\prime}}$ are
the potentials at the opposite (bottom and top, respectively) edges of the sample, and the
integral is taken across the sample from bottom to top. For non-local case the currents
are calculated from similar expressions:
\begin{eqnarray}
%7
I_{edge} =\frac{e^2}{h} \left(\varphi_{1}-\varphi_{2}+\varphi_{2^{\prime }}-\varphi_{1^{\prime }}\right),
\nonumber \\
I_{bulk}=\sum_{i=1,2} \left[ \sigma_{yx}^{\left( i\right) }\left( \psi_{i}-\psi_{i^{\prime}} \right)
-\sigma_{xx}^{\left( i\right) } \int dx \frac{\partial \psi_{i}}{\partial y} \right],
\end{eqnarray}
where now $\varphi_{i}$, $\psi_{i}$ and $\varphi_{i^{\prime }}$, $\psi_{i^{\prime}}$ are
the potentials at the opposite (left and right, respectively) edges of the current contact,
and the integral is taken across this current contact from left to right.

The conductivities are calculated as
$\sigma_{xx}^{(1)}=\sigma_{xx}^{(2)}=e (\mu_n n+\mu_p p)/2$, where
$\mu_n$ and $n$ ($\mu_p$ and $p$) are electron (hole) mobilities
and densities. To find $n$ and $p$, the bulk densities of states
for electrons and holes are represented by steps $D_{n,p} =4 \pi
m_{n,p}/h^{2}$ with $m_{n,p}$ being the effective masses of
electrons and holes. The energy gap separating electron and hole
bands is $E_{g}=3.3$ meV. In addition, the sharp band edges are
smoothed according to a Gaussian law with broadening energies
$\Gamma_{n,p}= \hbar/\tau_{n,p}$, where $\tau_{n,p}=\mu_{n,p}
m_{n,p}/e$. The following parameters have been used:
$\mu_{n}=80000$ cm$^{2}$/V s, $\mu_{p}=5000$ cm$^{2}$/V s,
$m_{n}=0.024~m_{0}$, $m_{p}=0.15~m_{0}$, where $m_0$ is the free
electron mass. The extended states are separated by a mobility gap
$\sim 1$ meV. Our calculations reproduce the temperature
dependence of the local and nonlocal resistances, shown in Fig. 2
of the main text. The best agreement between the experiment and
theory is reached for parameters $\gamma^{-1} = g^{-1}=5$ $\mu$m,
which roughly agrees with the mean free path $l$ represented in
the Table I of the main text. Parameters $\gamma$ and $g$ are
assumed to be independent of temperature.